\documentclass[a4paper,11pt]{article}
\usepackage{pos}
\usepackage{braket}
\usepackage{subfig}

\title{Pseudoscalar transition form factors and the hadronic light-by-light contribution to the muon $g-2$}
\ShortTitle{Pseudoscalar transition form factors and the HLbL contribution to the muon g-2}

\manuallySeparateAuthors
\author[a]{Antoine G\'erardin,}
\author[a,b]{ Jana N. Guenther,}
\author[a,b]{ Lukas Varnhorst}
\author{ and}
\author*[a]{ Willem E. A. Verplanke}
\author{ for the Budapest-Marseille-Wuppertal Collaboration}

\affiliation[a]{Aix-Marseille Universit\'e, Universit\'e de Toulon, CNRS, CPT,\\
  Marseille, France}

\affiliation[b]{Department of Physics, University of Wuppertal,\\
D-42119 Wuppertal, Germany}

\emailAdd{willem.verplanke@cpt.univ-mrs.fr}
\emailAdd{antoine.gerardin@cpt.univ-mrs.fr}

\abstract{We present preliminary results from our calculation of the pseudoscalar transition form factors of the $\eta$ and $\eta'$ mesons using staggered quarks on $N_f=2+1+1$ gauge ensembles generated by the Budapest-Marseille-Wuppertal collaboration. These transition form factors are an important input for the hadronic light-by-light contribution to the muon $(g-2)$. We first elaborate on the extraction of the masses of the $\eta$ and $\eta'$ mesons, that mix under the dynamics of QCD. Thereafter, we show our preliminary results for the pseudoscalar transition form factors, focusing on the $\eta$ meson in the absence of mixing.}

\FullConference{%
 The 38th International Symposium on Lattice Field Theory, LATTICE2021
  26th-30th July, 2021
  Zoom/Gather@Massachusetts Institute of Technology
}

\begin{document}
\maketitle

\section{Introduction}
Understanding the discrepancy between the theory calculation and the experimental value of the muon anomalous magnetic moment ($a_\mu$) \cite{Aoyama:2020ynm} is one of the most pressing challenges in present-day particle physics. The theoretical error is dominated by the hadronic vacuum polarization (HVP) and the hadronic light-by-light (HLbL) contributions. A recent computation of the leading order HVP by the Budapest-Marseille-Wuppertal (BMW) collaboration has however shed new light on the issue \cite{Borsanyi:2020mff}. In particular, it reduces the tension with the experimental value of $a_\mu$.

\begin{figure}[]
	\centering
	\includegraphics[width=0.72\textwidth]{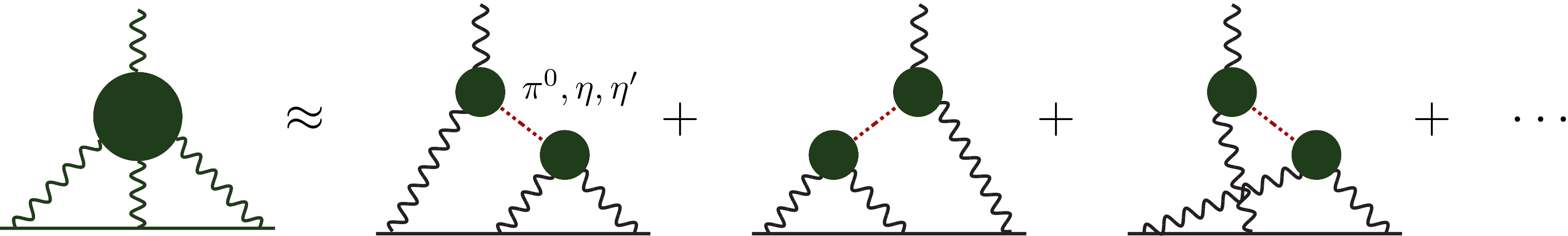}
	\caption{Hadronic light-by-light diagram and its decomposition into the dominant pseudoscalar poles. A wobbly line indicates a photon, the straight line the muon and a blob the non-perturbative hadronic interactions encoded in the pseudoscalar TFF.}
	\label{fig:hlbl}
\end{figure}

The HLbL interaction is shown in Figure \ref{fig:hlbl}. 
This contribution results from the scattering of four photons by means of a non-perturbartive hadronic interaction. Two model-independent routes to calculate the diagram are nowadays applied, namely the dispersive \cite{Colangelo:2014dfa, Colangelo:2014pva, Colangelo:2015ama} and lattice \cite{Chao:2021tvp, Blum_2020} computations. Crucial input for the dispersive approach are the transition form factors (TFFs) of the $\pi^0$, $\eta$ and $\eta'$ mesons, which are directly linked to pseudoscalar pole contributions to the $a_\mu^{\mathrm{HLbL}}$. The TFF describes the interaction of one on-shell pseudoscalar meson with two off-shell photons. The pion TFF has already been calculated \cite{Gerardin:2016cqj, Gerardin:2019vio}, and we provide here our first results towards a calculation of the $\eta$ and $\eta'$ TFFs. Note finally that, in experiment, the $\eta$ and $\eta'$ TFFs are difficult to access in the doubly virtual regime, where the photons have the same virtuality, so a theoretical calculation is of particular importance. ETM has already presented preliminary results of a calculation of the $\eta$, $\eta'$ TFFs \cite{burri2020}. The mixing between the two states encumbers the calculation, but only a 20\% precision on the contribution to $a_\mu^{\mathrm{HLbL}}$ is necessary to match future experimental precision \cite{Gerardin:2020gpp}.

In line with this interest, the second aim of this project is a calculation of the $\eta$ and $\eta'$ masses and mixing angles, using staggered quarks. Recent calculations of these quantities have been done using Wilson-Clover quarks \cite{bali2021masses} and twisted-mass quarks \cite{Ottnad:2017bjt}, and have shown good correspondence with experiment. The $\eta$ and $\eta'$ mesons are interesting particles to simulate, as the large mass of the $\eta'$ meson is directly related to the non-perturbative dynamics of QCD and the anomaly in the $U(1)_A$ charge symmetry of the QCD Lagrangian \cite{tHooft:1976snw}. Besides, flavor singlet mesons require the use of disconnected correlators, which are more challenging to compute and couple directly to the sea. The fourth-root trick employed for the sea quarks in staggered lattice QCD simulations has also been questioned \cite{Creutz:2007rk,Kronfeld:2007ek}. The $\eta'$ meson is hence of interest, and can therefore provide another hint towards the ability of staggered quarks to reproduce the axial anomaly in the continuum limit.
\section{Simulation details}
We use $N_f = 2+1+1$ dynamical staggered fermions with four steps of stout smearing generated by the BMW collaboration \cite{Borsanyi:2020mff}. These gauge ensembles are at nearly physical pion and kaon mass. We plan to exploit five different lattice spacings in the available range and consider $L=3$, $4$ and $6$ fm boxes for finite-size effect studies. Simulations are performed in the isospin limit where $m_u = m_d \equiv m_\ell$. In Table \ref{tab:ensembles} we summarize the details of the two ensembles used in this preliminary study. The three-point functions have only been computed on the coarse ensemble.
\begin{table}
	\centering
	\begin{tabular}{p{2cm}|p{1cm}| p{1cm} | p{1.6cm} | p{1cm}}
		&$\beta$& a[fm] & $L/a\times T/a$ & \# conf\\
		\hline
		three-point&3.7000& 0.1315 & 32 $\times$ 64 & 900\\
		two-point&3.8400& 0.0952 & 32 $\times$ 64 & 1100
	\end{tabular}
	\caption{Summary of two ensembles with lattice spacing, lattice size and number of gauge configurations.}
	\label{tab:ensembles}
\end{table}
\section{$\eta$ and $\eta'$ mass}
\subsection{Mass extraction}
The analysis of the $\eta-\eta'$ system starts by the quark model. One introduces an SU(3) octet and singlet field
\begin{align}
	\eta_8(x) &= \frac{1}{\sqrt{6}}\left(\overline{u}\gamma_5 u(x) + \overline{d}\gamma_5 d(x) - 2\overline{s}\gamma_5 s(x)\right),\\
	\eta_0(x) &= \frac{1}{\sqrt{3}} \left(\overline{u}\gamma_5 u(x) + \overline{d}\gamma_5 d(x) + \overline{s}\gamma_5 s(x)\right).
\end{align}
The $\eta_8(x)$ field is a Goldstone realization of spontaneously broken symmetry group $G_F = U(1)_V \otimes SU(3)_L \otimes SU(3)_R \overset{SSB}{\longrightarrow} U(1)_V \otimes SU(3)_V$ of the classical QCD Lagrangian. On the other hand, $\eta_0$ is the would-be Goldstone boson realization of the $U(1)_A$ charge symmetry, that is broken at the quantum level. Both fields have quantum numbers $J^{PC} = 0^{-+}$, like the $\pi^0$. However, no mixing with the pion occurs in our lattice simulations since we work in the isospin limit.

We consider the matrix of correlators built from the lattice operators $O_8$, $O_0$ that interpolate the SU(3) fields
\begin{align}
	\mathcal{C}(t) &\equiv \begin{pmatrix}
		\langle O_8(t)O_8^\dagger(0)\rangle & \langle O_8(t)O_0^\dagger(0)\rangle \\ \langle O_0(t)O_8^\dagger(0)\rangle& \langle O_0(t)O_0^\dagger(0)\rangle \end{pmatrix}\nonumber \\
	&= \begin{pmatrix}
		\frac{1}{3}\left(C_\ell + 2C_s - 4D_{\ell s} + 2D_{\ell \ell} + 2D_{ss}\right)&\frac{\sqrt{2}}{3}\left(C_\ell - D_{\ell s} - D_{ss}-C_s + 2D_{\ell\ell}\right)\\ \frac{\sqrt{2}}{3}\left(C_\ell-D_{\ell s} - D_{ss}-C_s + 2D_{\ell\ell}\right)&\frac{1}{3}\left(2C_\ell+C_s + 4D_{\ell\ell} + 4D_{\ell s} + D_{ss}\right)
	\end{pmatrix},
\label{eq:gevp}
\end{align}
where $C_q$ and $D_{qq'}$ are respectively the connected and disconnected correlator for quark flavor $q,q' = \ell,s$ and we have made use of $D_{ls} = D_{sl}$. Notice that the non-vanishing off-diagonal terms explicit the mixing between $\eta_8$ and $\eta_0$; furthermore, even in the $SU(3)_F$ limit, the disconnected contributions to $\langle O_0(t)O_0^\dagger(0)\rangle$ do not vanish. Lastly, in this chiral limit, the physical $\eta$ and $\eta'$ state correspond exactly to the $\eta_8$ and $\eta_0$ state.

Masses of the $\eta$ and $\eta'$ mesons can then be obtained by solving a Generalized Eigenvalue Problem (GEVP) \cite{collaboration_2009}
\begin{equation}
	\mathcal{C}(t)v_n(t,t_0) = \lambda_n(t,t_0) \mathcal{C}(t_0) v_n(t,t_0).
\end{equation}
Here $\mathcal{C}(t)$ is an $N\times N$ matrix, $\lambda_n$ are the eigenvalues, $v_n$ are the eigenvectors, $t_0/a$ is a free parameter that is fixed to $1$ in this analysis, and $n \in \{1,\dots,N\}$. From the eigenvalues one can define an effective mass that converges to the meson mass at large time
\begin{equation}
	E_n^{\textrm{eff}}(t) = \log\left(\frac{\lambda_n(t,t_0)}{\lambda_n(t+1,t_0)}\right) \overset{t\to\infty}{\longrightarrow} m_n,
\end{equation}
in the region where the backward propagating quarks can be neglected. In the next section we present the lattice operators that interpolate the $\eta$, $\eta'$ states in the staggered quark formalism.
\subsection{Staggered mesonic operators}
Staggered mesonic operators have been classified by Golterman in \cite{Golterman:1985dz}. Following the notation of \cite{Altmeyer:1992dd}, there are two taste-singlet operators that couple directly to the $\eta$, $\eta'$ mesons
\begin{align}
	\mathcal{O}_3(x) &= \frac{1}{6}\sum_{ijk}\epsilon_{ijk}\overline{\chi}(x)\left[\eta_i\Delta_i\left[\eta_j\Delta_j\left[\eta_k\Delta_k\right]\right]\right]\chi(x) \equiv \overline{\chi}(x) \hat{O}_3\chi(x),\\
	\mathcal{O}_4(x) &= \frac{1}{2}\eta_4(x)\left[\overline{\chi}(x)\hat{O}_3\chi_+(x) + \overline{\chi}_+(x)\hat{O}_3\chi(x)\right].
\end{align}
Here $\chi(x)$ is a fermionic field and $\epsilon_{ijk}$ is the Levi-Civita tensor; the symmetric shift $\Delta_\mu$, $\chi_+(x)$  and the staggered phase factors $\eta_\mu(n)$ in the convention $n=(x,y,z,t)$ are
\begin{align}
	\Delta_\mu\chi(x) &= \frac{1}{2}\left[U_\mu(x)\chi(x+\hat{\mu}) + U_\mu^\dagger(x-\hat{\mu})\chi(x-\hat{\mu})\right],\\
	\chi_+(x) &= U_4(x)\chi(x+\hat{t}), \quad \overline{\chi}_+(x) = \overline{\chi}(x+\hat{t})U_4^\dagger(x),\\
	\eta_\mu(n) &= (-1)^{\sum_{\nu<\mu}n_\mu}.
\end{align}
In the ubiquitous staggered operator notation $\Gamma_D\otimes\Gamma_F$, where $\Gamma_D$ denotes the Dirac structure, and $\Gamma_F$ the taste structure, the 3-link operator $\mathcal{O}_3(x)$ couples to an axial vector $\gamma_4\gamma_5 \otimes 1$ and a parity partner scalar meson $1\otimes \gamma_4\gamma_5$. The parity partner for the 3-link operator creates an oscillatory behavior in the effective energy for the taste-singlet pion. This makes a trustworthy extraction of the pseudoscalar ground state energies  cumbersome. The 4-link operator $\mathcal{O}_4(x)$, which is non-local in time, couples to a pseudoscalar $\gamma_5\otimes 1$ and an exotic state $\gamma_4 \otimes \gamma_4 \gamma_5$. As noted in  \cite{Altmeyer:1992dd, Kilcup:1986dg}, the coupling to the parity partner state for such an operator is strongly suppressed. Hence, it is our choice of mesonic operator throughout the analysis, also because the separation by an even amount of links allows for a computational trick, as is discussed next.
\subsection{Pseudoscalar two-point functions}
\begin{figure}
	\centering
	\includegraphics[width=0.75\textwidth]{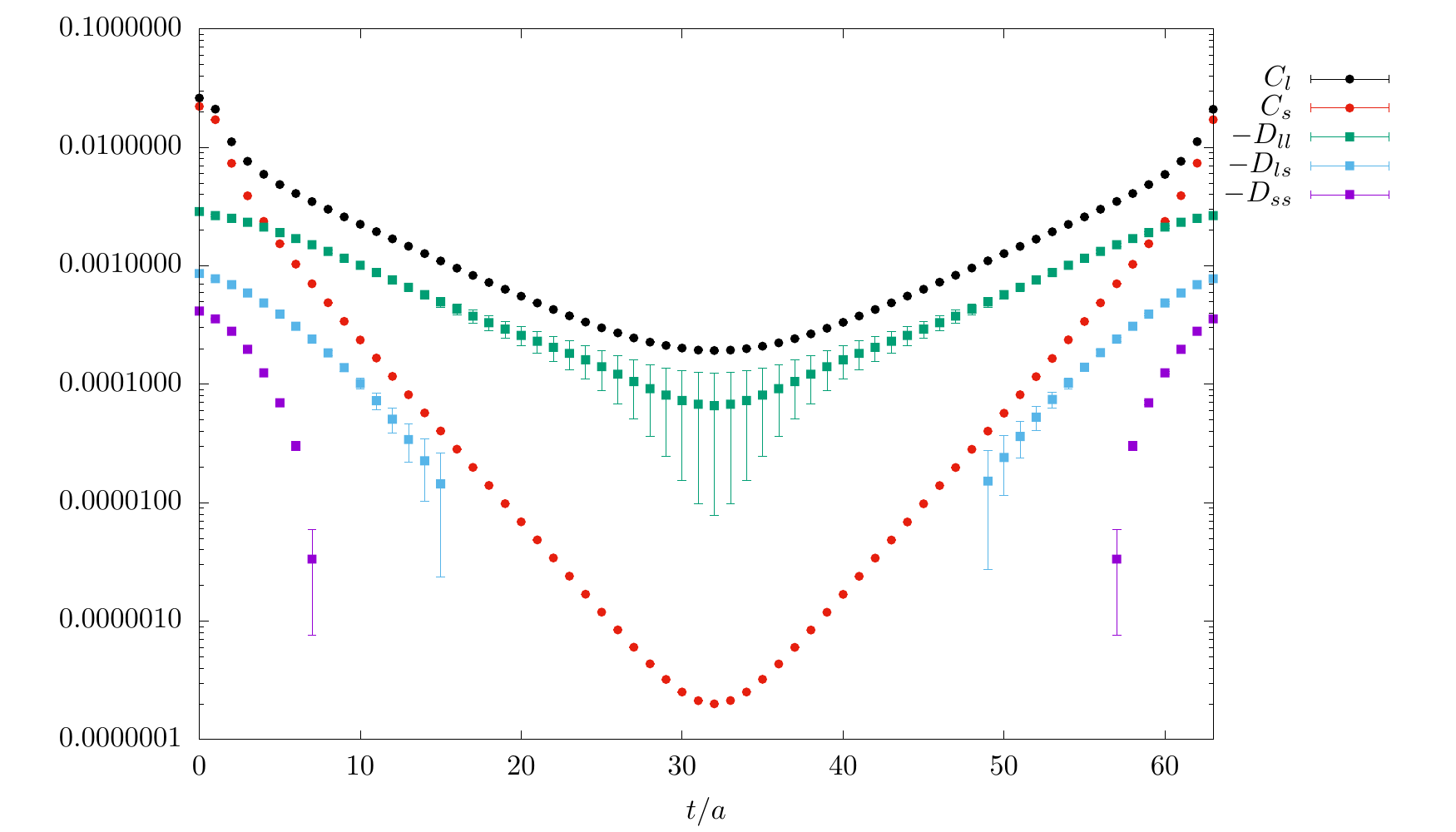}
	\caption{Different connected $C_q$ and disconnected $D_{qq'}$ correlation functions contributing to the $\eta$, $\eta'$ spectroscopic analysis.}
	\label{fig:correlators}
\end{figure}

We apply several noise-reduction tricks to compute the correlation functions effectively. Firstly, we use low-mode averaging (LMA) \cite{Giusti_2004, DeGrand_2005} with $n=300$ modes in our $4$ fm box (and scale modes with the volume of the box). Secondly, we apply all-mode averaging (AMA) \cite{Bali_2010, Blum_2013} for the stochastic part of our estimator. And lastly, we employ a Venkataraman-Kilcup reduction trick \cite{Venkataraman:1997xi} for our one-point functions. In this context, it uses the staggered Dirac operator that connects only even and odd sites, together with the 4-link operator which only couples even (odd) and even (odd) sites. This allows one to construct a different pseudoscalar loop estimator that has a decreased variance. As a consequence of all these tricks, we have managed to reach the gauge noise for the two-point disconnected correlation functions. In Figure \ref{fig:correlators}, we show the result for all our correlation functions. Note that we find a good signal for $C_\ell$, $C_s$ and $D_{\ell\ell}$, while the signal for the $D_{\ell s}$ and especially the $D_{ss}$ are lost in the noise quickly. This is however not an issue, since the effect of the $D_{ss}$ correlator on the GEVP is extremely small, and to a lesser extent this holds for the $D_{\ell s}$ as well.
\subsection{Analysis techniques}
\label{ssec:analysistechniques}
\begin{figure}[]
	\centering
	\subfloat[]{\includegraphics[width=0.33\textwidth]{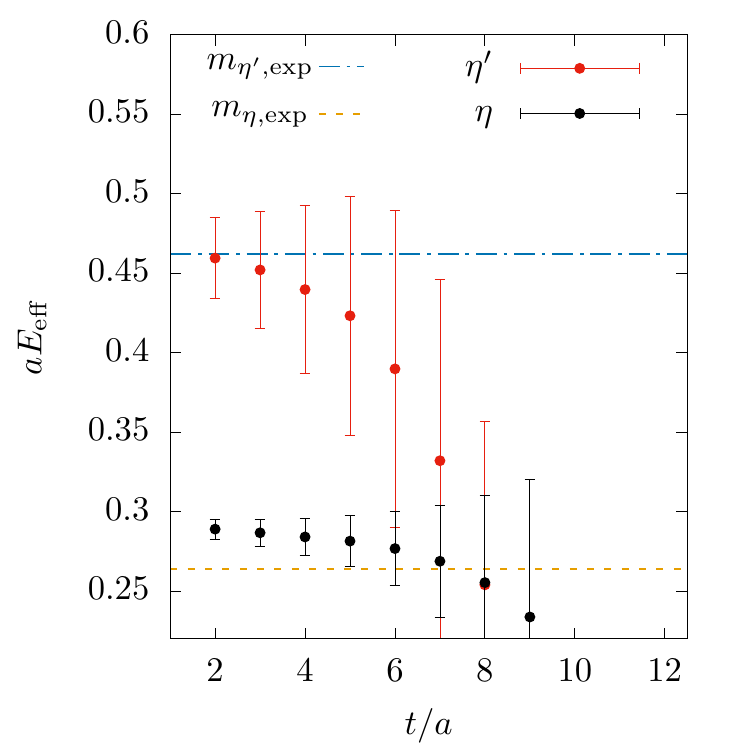}\label{fig:eff_mass_prel}}
	\hfill
	\subfloat[]{\includegraphics[width=0.33\textwidth]{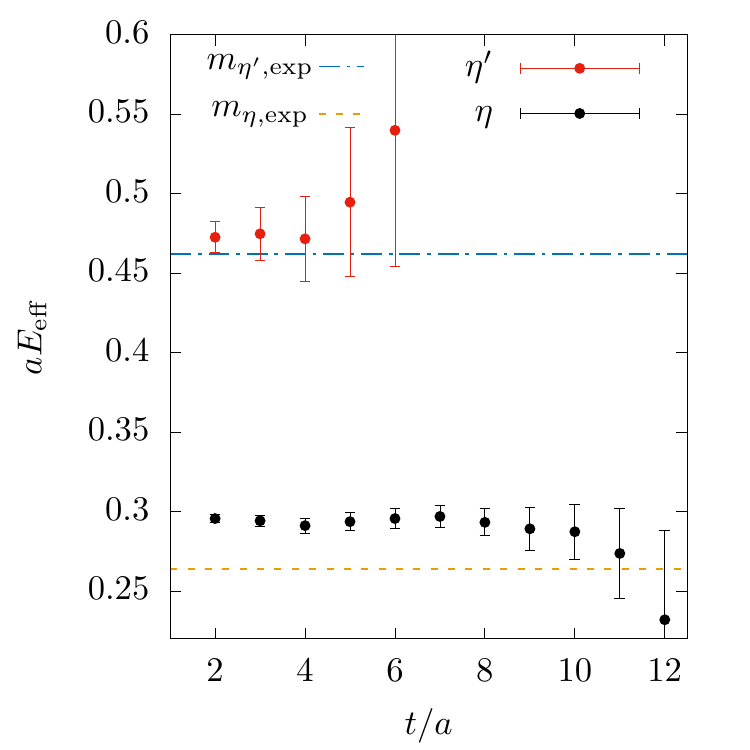}\label{fig:eff_mass_prel2}}
	\hfill
	\subfloat[]{\includegraphics[width=0.33\textwidth]{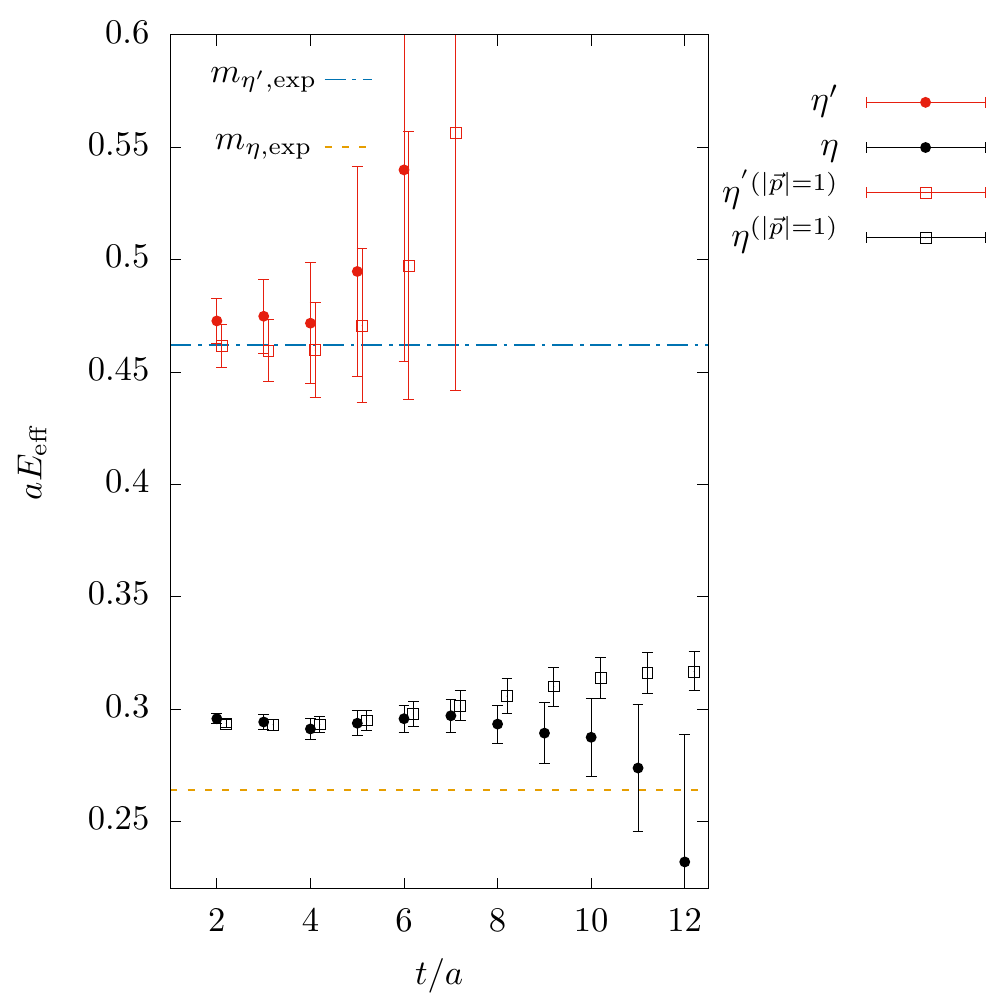}\label{fig:eff_mass_full}}
	\caption{Effective mass plots for three scenarios. Left: after applying method $1$ of section \ref{ssec:analysistechniques}; Middle: Additionally applied method (2); Right: Added the $|\vec{p}|=1\cdot\left(\frac{2\pi}{L}\right)$ result. Dashed lines indicate the experimental values of the $\eta$ and $\eta'$ masses.}
	\label{fig:effmass}
\end{figure}

Extracting the mass of the $\eta$ and $\eta'$ mesons is hindered by the exponential growth of the noise/signal ratio. In fact, by only using a GEVP, the signal of the two mesons is lost before reaching a plateau. Thus, to to circumvent this issue, we apply two analysis techniques. Firstly, we employ a fitting trick to remove excited state contributions in the connected correlator, which was first introduced in this context by ETM \cite{Michael_2013}. The underlying assumption is that most excited state contamination is found in the connected correlation functions. Since these have a relatively good signal over noise ratio (however not constant, as we work with taste-singlet quantities), it is possible to reliably determine the ground state of these correlators. Connected mesonic correlation functions calculated on a lattice with periodic boundary conditions, assume a spectral decomposition of the form
\begin{equation}
	C_q(t) = A_q \left(\exp(-E_q^{(1)} t) + \exp(-E_q^{(1)}(T-t))\right) + \mathcal{O}(\exp(-E_q^{(2)}t)), \quad q=\ell,s.
\end{equation}
Fitting $C_q$ to this function in a region where excited states are negligible, one can safely determine the energy $E_q$ and overlap $A_q$ of the ground state. Then one replaces the connected contribution in equation (\ref{eq:gevp}) by this one-exponential fit on the whole time range, leaving the disconnected contributions unchanged. If excited state contributions in the disconnected correlators are indeed small, excited states should effectively be removed and a plateau for the effective mass should be reached early in time. Results of applying this method to our data are shown in Figure \ref{fig:eff_mass_prel}. We cannot resolve excited states at this stage, but errors are still large, especially for the $\eta'$ effective mass.

The second trick was first introduced in \cite{Ottnad:2017bjt}, and results from an analytic understanding of correlation functions in finite volume \cite{Aoki_2007}. In fact, for $\vec{p}=\vec{0}$, disconnected correlation functions in finite volume, do not tend to zero at large time if the topological charge $Q$ isn't correctly sampled. One finds that
\begin{align}
	D_{qq}(t) \overset{t\to\infty}{\sim} \frac{a^5}{T}\left(\chi_t - \frac{Q^2}{V}+ \frac{c_4}{2V\chi_t}\right).
\end{align}
Here, $\chi_t$ is the topological susceptibility and $c_4$ is the kurtosis of the topological charge. Such a constant shift can be removed by taking a discrete derivative of the matrix of correlators $\mathcal{C}(t)$,
\begin{align}
\mathcal{C}(t) \rightarrow \mathcal{C}'(t) \equiv \mathcal{C}(t) - \mathcal{C}(t+ \Delta t).
\end{align}
This shifted correlation matrix still satisfies a GEVP with $\Delta t/a$ as a free parameter. Aside of removing this constant shift, it also helps reducing correlations between time-slices and hence improves the point error. In Figure \ref{fig:eff_mass_prel2} we show the result of applying this method to our data, alongside with the first trick, and see that now a plateau is found for $\eta$ and $\eta'$ and errors are reduced drastically. Further, in Figure \ref{fig:eff_mass_full} we show the result for our correlation functions which carry one unit of momentum. The errors are slightly smaller and the data are compatible with $\vec{p}=\vec{0}$, motivating a use of both kinematic frames in future analyses. Lastly, we observe a mild discretization effect on the $\eta'$ meson, while it's of the order of 10\% for the $\eta$ meson.

\section{Pseudoscalar transition form factors}
\begin{figure}[h]
	\centering
	\subfloat[]{\includegraphics[width=0.65\textwidth]{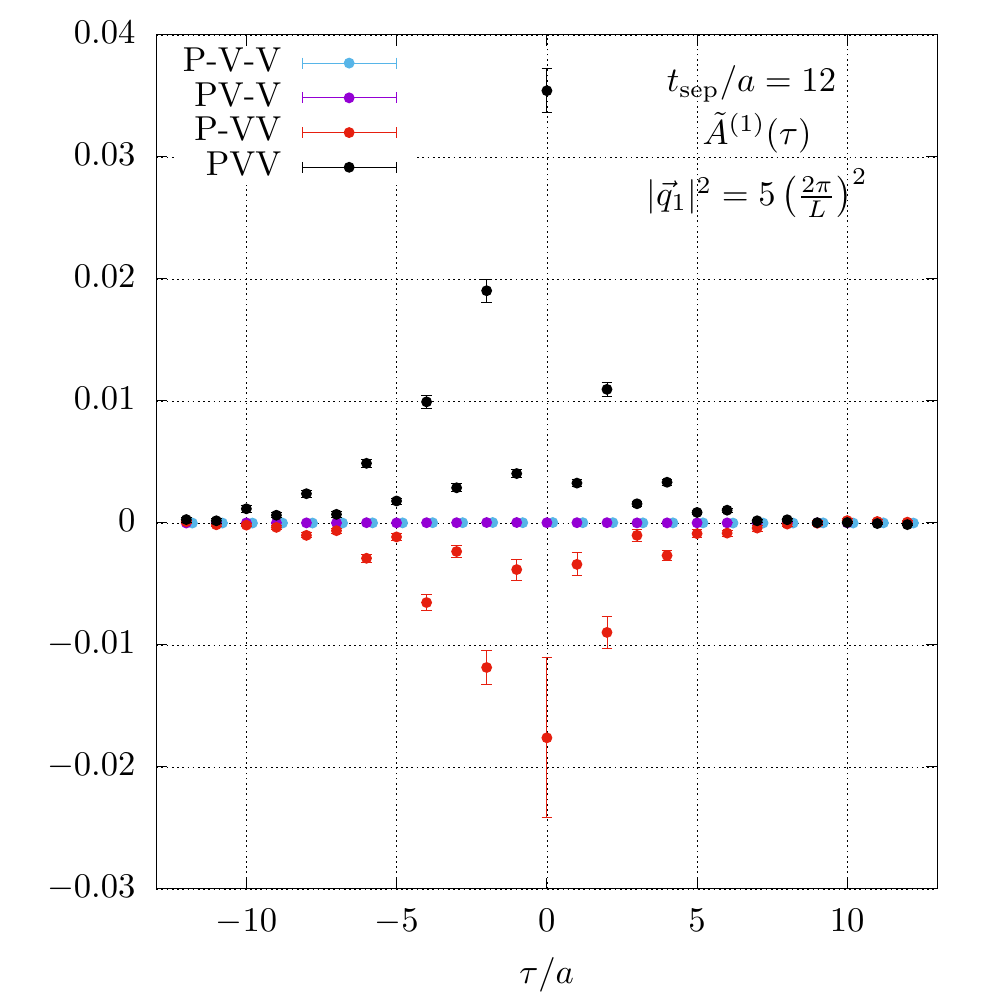}\label{fig:integrand}}
	\hfill
	\subfloat[]{\includegraphics[width=0.3\textwidth]{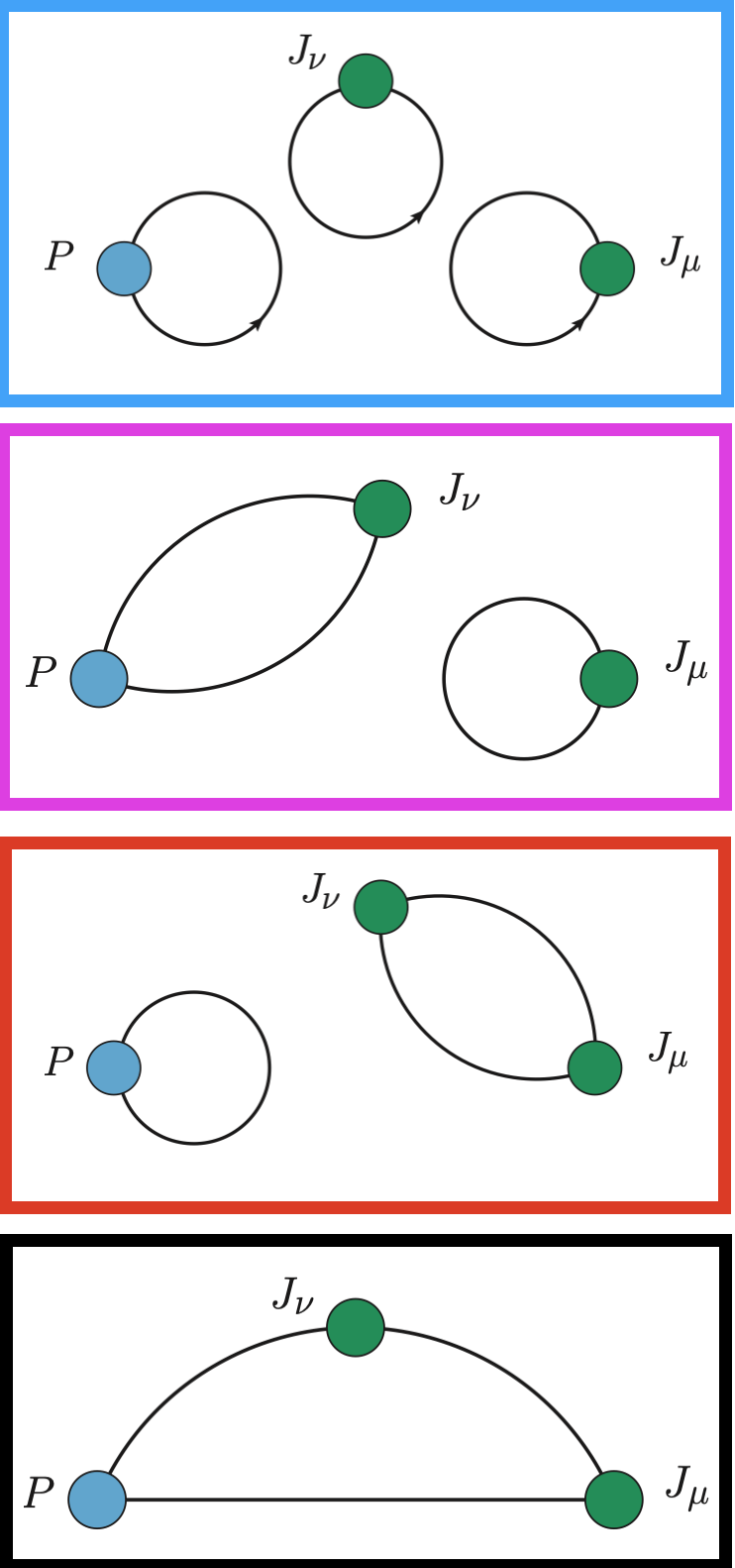}\label{fig:topologies}}
	\caption{Left: The integrand of the $\eta_8$ TFF for specified kinematic condition; $t_{sep}$ refers to the seperation between the pseudoscalar and one of the vector currents that is kept fixed. Right: Illustration of the different topologies of the connected and disconnected contributions to the three-point function.}
	\label{fig:int+top}
\end{figure}
In Minkowski space-time, the TFF for a pseudoscalar meson $\mathcal{F}_{\mathrm{p}\gamma^*\gamma^*}(q_1^2,q_2^2)$ is defined by the following matrix elements $M_{\mu\nu}$ 
\begin{equation}
	M_{\mu\nu}(p,q_1) = i\int d^4x\, e^{iq_1\cdot x}\bra{\Omega}T\{J_\mu(x)J_\nu(0)\}\ket{P(p)}=\epsilon_{\mu\nu\alpha\beta} \,q_1^\alpha q_2^\beta \mathcal{F}_{\mathrm{p}\gamma^*\gamma^*}(q_1^2,q_2^2), 
\end{equation}
where $q_1$ and $q_2$ are the photon $4$-momenta, $J_\mu$ is the hadronic component of the electromagnetic current and $\epsilon_{\mu\nu\alpha\beta}$ is a $4$-rank Levi Civita tensor. On the lattice, one can relate three-point functions to the Euclidean version of these matrix elements \cite{Ji_2001}. We write 
	\begin{equation*}
	M_{\mu\nu}^E = \frac{2E_{P}}{Z_{P}}\int_{-\infty}^{\infty}d\tau\, e^{\omega_1\tau}\tilde{A}_{\mu\nu}(\tau),
\end{equation*}
where $E_{P}$ and $Z_{P}$ are respectively the pseudoscalar energy and overlap of the pseudoscalar state with the interpolating operator, $\omega_1$ is a free parameter $q_1 = (\omega_1, \vec{q}_1)$, $q_2 = (E_P - \omega_1, \vec{q}_2)$, and 
\begin{align*}
	\tilde{A}_{\mu\nu}(\tau) &\equiv \lim\limits_{t_{P}\to\infty} e^{E_{P}(t_f-t_0)} C_{\mu\nu}^{(3)}(\tau,t_{P}), \\
	C_{\mu\nu}^{(3)}(\tau,t_{P}) &= a^6 \sum_{\vec{x},\vec{z}} \langle J_\mu (\vec{z},t_i)J_\nu(\vec{0},t_f)P^\dagger(\vec{x},t_0)\rangle e^{i\vec{p}\cdot \vec{x}}e^{-i\vec{q}_1\cdot \vec{z}}.
\end{align*}
Here we have defined $\tau = t_i -t_f$, the time-difference between the two vector currents and $t_{P}=\mathrm{min}(t_f-t_0,t_i-t_0)$, the minimal time separation between the pseudoscalar and the vector currents. In the three-point function, $P$ is the interpolating operator for the pseudoscalar meson, expressed in terms of $O_8$, $O_0$ and appropriate mixing parameters; $J_\mu$ is implemented using the conserved vector current which does not require renormalization \cite{Borsanyi:2020mff}. Note also that, an accurate determination of the mixing parameters and the masses of the $\eta$, $\eta'$ mesons is important, as they enter in the three-point function.

$\tilde{A}_{\mu\nu}$ can  be decomposed into two scalar functions $\tilde{A}^{(1)}$, $\tilde{A}^{(2)}$ that form the integrand of the TFF (for technical details we refer to \cite{Gerardin:2019vio}). In Figure \ref{fig:int+top}, we show preliminary results for one of these scalar functions in the $\vec{p}=\vec{0}$ frame, where we focus on $\eta_8$, thus discarding any mixing with $\eta_0$. The latter is important, and will be implemented in the future. Note that four different topologies contribute to the integrand (the hyphen indicates a disconnection): the two dominant PVV connected and P-VV disconnected loops; and the negligible PV-V and P-V-V disconnected loops. The P-VV loop has a significant and negative contribution and therefore we observe a large cancellation with the PVV loop. As a consequence, a good control of the disconnected VV and P loops is important.

\section{Conclusions}
We have shown preliminary results toward the extraction of the $\eta,\eta'$ mass using staggered quarks. Using the GEVP and several noise reduction techniques, we are able to obtain a good signal for the $\eta$ and $\eta'$ effective mass. Alongside this, we have shown promising preliminary results for the integrand of the TFF of the $\eta_8$ state. In this context we have found that the two leading contributions are the fully connected PVV and disconnected P-VV loops, while the disconnected PV-V and P-V-V loops are of a marginal size. 

Presently, we are accumulating statistics for our P-VV contribution, to further reduce its error. At the same time, we will explore the signal of the $\eta$, $\eta'$ and their TFFs, including therefore mixing between the states. We plan to do this on several lattice spacings and volumes, improving also the statistics on the different ensembles that we have already analyzed. Ultimately, we will perform a continuum extrapolation of the TFFs.
\section{Acknowledgements}
This publication received funding from the Excellence Initiative of Aix-Marseille University - A*MIDEX, a French “Investissements d’Avenir” programme, AMX-18-ACE-005 and from the French National Research Agency under the contract ANR-20-CE31-0016. Center de Calcul Intensif d'Aix-Marseille is acknowledged for granting access to its high performance computing resources. 

\bibliographystyle{JHEP}
\begingroup
\fontsize{10pt}{3pt}\selectfont
\bibliography{bibliography}
\endgroup
\end{document}